\documentclass[conference, compsoc]{IEEEtran}

\usepackage{filecontents}
\usepackage{cite}
\usepackage{amsmath,amssymb,amsfonts}
\usepackage{graphicx}
\usepackage{textcomp}
\usepackage{xcolor}
\usepackage{algorithm}
\usepackage{algorithmicx}
\usepackage{threeparttable}
\usepackage{threeparttablex}
\usepackage{algpseudocode}
\usepackage{amsmath}
\usepackage{listings}
\usepackage{float}
\usepackage{url}
\usepackage{balance}
\usepackage{booktabs}
\usepackage{tabularx}
\usepackage{longtable}
\usepackage{array}
\usepackage{multicol}
\usepackage{multirow}
\usepackage{adjustbox}
\usepackage{tikz}
 \usepackage{xspace}
\usetikzlibrary{positioning, calc, shapes}
\newcommand{\tool}{\textsc{RECON}\xspace}

\usepackage[many]{tcolorbox}
\newtcolorbox[auto counter]{mybox}[2][]{enhanced jigsaw, breakable, #1}

\definecolor{junkbyte}{RGB}{255,200,200}  % Pink for junk bytes
\definecolor{patterncolor}{RGB}{255,255,150} % Yellow for pattern
\definecolor{skipcolor}{RGB}{144,238,144} % Lime/Green for SKIP target

\usepackage{xcolor}

\lstdefinestyle{jimple}{
    language=Java,
    basicstyle=\footnotesize\ttfamily,
    numbers=left,
    numberstyle=\tiny\color{gray},
    frame=single,
    captionpos=b,
    breaklines=true,
    showstringspaces=false,
    tabsize=2,
    keywordstyle=\color{blue},
    commentstyle=\color{green!50!black},
    stringstyle=\color{red},
    escapeinside={(*@}{@*)},
    morekeywords={specialinvoke,virtualinvoke,interfaceinvoke,goto,if,label}
}
\algrenewcommand\algorithmicrequire{\textbf{Input:}}
\algrenewcommand\algorithmicensure{\textbf{Output:}}

\begin{document}
%-------------------------------------------------------------------------------

% Don't want date printed
\date{}

% Title
\title{\Large \bf \tool: An LLM-Enhanced Backward Constraint Analysis Framework }

% Anonymous authors for submission
\author{
{\rm Babangida Bappah, Lamine Noureddine, 
Umar Farooq, Aisha Ali-Gombe}\\
Louisiana State University\\
\{bbappa1, lnoureddine, ufarooq, aaligombe\}@lsu.edu
}

\maketitle

%-------------------------------------------------------------------------------
\begin{abstract}
%-------------------------------------------------------------------------------
While traditional techniques, such as symbolic execution, provide a principled foundation for precise constraint reasoning in program analysis, they struggle to scale to modern software systems mainly due to path explosion, the need for function modeling, and the loss of semantic intent at low-level program representations. In complex execution environments such as Android, characterized by extensive framework interactions and event-driven behavior, these limitations are even more amplified. Thus, in this paper, we present a novel large language model (LLM)-enhanced backward constraint analysis framework that combines the precision of static program analysis with LLM's semantic understanding to extract precise execution constraints from Android bytecode. Our approach, titled RECON, performs backward path discovery from target method(s) to the application entry point(s), discovers method-level control-flow constraints, and leverages LLM reasoning to transform bytecode conditions into interpretable specifications. We evaluated RECON using five LLMs across 78 Android constraint-extraction scenarios and compared it with traditional symbolic execution on real-world applications. Results demonstrate that our approach operates 5.8X faster than traditional symbolic execution, with a 100\% success rate, while maintaining logical equivalence and providing significantly more precise and interpretable output. We further evaluated RECON for malware analysis on 100 samples. The results indicate an 84\% success rate in generating semantic constraints that lead to the execution of dangerous API behaviors and in detecting complex constraints across multiple execution paths.

%The results show an 84\% success rate in generating semantic constraints that lead to the execution of dangerous API behaviors. Additionally, our evaluation reveals that \tools can detect complex constraints, enabling security analysts to understand malicious behavior without extensive manual reverse engineering. 

%geared towards privacy/malware analysis and security verification, 
\end{abstract}

%-------------------------------------------------------------------------------
% Main content sections
%-------------------------------------------------------------------------------
\section{Introduction}

Understanding the conditions under which sensitive operations become reachable in large, framework-intensive applications remains a fundamental challenge in program analysis. Traditional symbolic execution is widely used for constraint analysis and discovering program behaviors. Yet, its scalability is affected by path explosion and its accuracy is constrained by the need for precise function models—two problems that are particularly severe in object-oriented, event-driven, and framework-based programs. Current engines such as KLEE~\cite{cadar2021klee}, JPF\cite{anand2007jpf}, Soot\cite{vallee1999soot}, angr\cite{shoshitaishvili2016state}, SymCC\cite{255310}, Frama-C\cite{cuoq2012frama} either punt on modeling, rely on hand-written stubs, or explode on heap / OO programs. They struggle to determine the conditions under which specific program contexts or target methods are reachable, especially in real-world applications where framework logic, callbacks, and library code obscure the true control-flow structure. In modern Android applications, particularly complex control flow, extensive framework interactions, and event-driven execution obscure the execution prerequisites of sensitive behaviors, making manual analysis costly and traditional symbolic techniques difficult to scale. While static analysis provides broad coverage and dynamic analysis offers high fidelity, both struggle to recover interpretable execution constraints in real-world applications. At the same time, large language models (LLMs) possess unprecedented ability to infer high-level semantics of methods, classes, and API behavior. Yet, these capabilities remain largely unused in constraint analysis pipelines. This creates a fundamental research gap: how can we design a constraint reasoning framework that leverages LLM semantic insight to infer method behavior, discover constraints along targeted call paths, and assemble them into a feasible global path condition—without exhaustive exploration or manual modeling.

Thus, in this work, we present \tool, a target-driven constraint analysis framework that accurately recovers human-interpretable reachability requirements for target operations in Android applications. \tool combines backward call-graph discovery, upward intraprocedural constraint analysis, and LLM-assisted semantic interpretation to transform low-level bytecode conditions into execution requirements. Rather than constructing fully formal symbolic path formulas, our approach focuses on recovering the necessary reachability conditions suitable for scalable security analysis and human interpretation.

Our evaluation results show that the LLM-assistant semantic extraction integrated with GPT-4o in \tool consistently achieves 92.6\% semantic interpretation accuracy across Android framework APIs and demonstrates constraint recovery precision of 88.2\% even in complex control-flow scenarios, significantly outperforming open-source LLM alternatives. Our comparative analysis against angr\cite{shoshitaishvili2016state} shows that \tool recovers logically equivalent reachability conditions when both analyses complete, while achieving markedly lower analysis time and higher completion rates (5.8X faster) on real-world applications. Additionally, evaluation on Android malware demonstrated the framework’s ability to identify and explain execution prerequisites for security-sensitive behaviors across diverse malware families, translating complex bytecode dependencies into human-interpretable requirements. Collectively, these results validate \tool as a practical and scalable approach for understanding execution reachability in modern Android applications, particularly in security analysis contexts where finding and interpreting the necessary conditions for a sensitive operation to occur are critical, rather than exhaustive symbolic search.

This paper makes the following contributions:
\begin{itemize}
\item We present a constraint reasoning framework that combines static program analysis with LLM–assisted semantic reasoning to interpret bytecode-level conditions that are difficult to reason using traditional symbolic analysis.
\item We introduce a scalable framework design that employs target-driven backward call-graph discovery, upward intraprocedural constraint reasoning, and lightweight path-level constraint assembly to efficiently recover reachability requirements in large, event-driven Android applications.
\item Our approach accurately recovers human-interpretable reachability constraints that explain the execution prerequisites of target operations, enabling practical use in security analysis.
\item We evaluate \tool on benign real-world applications and Android malware, demonstrating its practical utility, scalability, and effectiveness in recovering execution constraints across diverse analysis scenarios.     
\end{itemize}

\section{Background}
Constraint analysis evaluates the conditions that must hold for a program to reach specific execution points or produce particular behaviors. These conditions, commonly expressed as logical predicates over program variables, are typically derived from control-flow decisions, data dependencies, and interactions with external libraries or system components. Constraint-based reasoning has therefore served as a foundational technique for a wide range of software engineering tasks, including testing, verification, vulnerability discovery, and security analysis~\cite{gulwani2008program, cadar2011symbolic, puasuareanu2019symbolic}.
Many such program analysis tasks can be formulated as constraint-solving problems, in which verification conditions, invariants, or program properties are encoded as logical constraints and discharged using SAT or SMT solvers~\cite{gulwani2008program}. By solving these constraints, analysis frameworks can automatically reason about program behavior, enabling capabilities such as test generation, detection of unreachable states, and identification of security-sensitive behaviors~\cite{luo2017system}.

\subsection{Techniques for Constraint Analysis}

\paragraph{Static Techniques: }Static constraint analysis techniques derive execution conditions from program structure without executing the program. These approaches rely on symbolic reasoning, abstract interpretation, data-flow analysis, and constraint encoding to approximate program behavior across feasible execution paths. Symbolic execution is one of the most widely adopted static analysis techniques for constraint analysis. It construct path constraints by representing inputs symbolically and reasoning over control-flow decisions, enabling automated analysis of reachability, test generation, and verification properties \cite{cadar2011symbolic,puasuareanu2013symbolic,puasuareanu2019symbolic}. Abstract interpretation is another static analysis technique for creating approximate semantics of programs, allowing for sound reasoning about their dynamic behaviors\cite{cousot1996abstract,cousot1977abstract,cousot2014abstract,giacobazzi2024adversities}. CEGAR on the other hand is lazy approach that start with a coarse abstraction and refining only when needed\cite{clarke2000counterexample}. In general, static analysis provides broad behavioral coverage by reasoning about paths that may be difficult or infeasible to observe during runtime testing. Many static techniques also provide conservative guarantees through over-approximation, which is particularly valuable for security verification and policy enforcement. Since static analysis does not require program execution, it can be applied to incomplete programs, unsafe binaries, or unavailable runtime environments. 

\paragraph{Dynamic Techniques: }Dynamic constraint analysis techniques, on the other hand, infer execution constraints from observed program behavior during concrete program execution. These methods commonly combine runtime execution tracing with symbolic reasoning, such as concolic execution, which explores alternative execution paths by collecting path constraints during execution and generating new inputs to trigger unexplored behaviors~\cite{godefroid2005dart, sen2006cute}. Modern symbolic execution engines extend this paradigm by combining runtime exploration with solver-based path-feasibility checking and model-checking techniques \cite{puasuareanu2013symbolic}. Other techniques, such as taint analysis, dynamically monitor the data flow from source to sink, often optimizing concolic analysis by concentrating symbolic computation solely on tainted variables\cite{clause2007dytan,newsome2005dynamic}. Dynamic execution trace leverages instrumentation to monitor user/program/system interaction at runtime\cite{hojaji2019model,larus1990abstract}, while invariant mining techniques explore the inherent linear characteristics of
runtime program workflows, which are often used for detecting anomalies\cite{lou2010mining,cheng2008simulation,lemieux2015mining}. In general, dynamic approaches provide high precision because constraints are derived from real execution states, naturally capturing interactions with the environment and runtime behaviors that are difficult to model statically. This makes dynamic techniques particularly effective in event-driven systems and security testing scenarios where accurate modeling of runtime behavior is critical ~\cite{gao2018android, luo2017system}. As a result, dynamic constraint analysis is widely used in vulnerability discovery, exploit development, and malware analysis.    

\subsection{Challenges of Existing Constraint Analysis}
Despite significant progress in constraint-based program reasoning, existing techniques face well-known limitations that restrict their scalability, precision, and applicability to modern software systems. Modern software systems exhibit complex control flows, extensive framework interactions, and large input spaces, all of which substantially increase the difficulty of accurately discovering and solving execution constraints~\cite{puasuareanu2019symbolic, bu2021machine}. Static constraint analysis techniques, in particular, often suffer from precision loss due to necessary over-approximation of program behavior. To preserve soundness, static analyses frequently model multiple potential execution paths simultaneously, which can introduce spurious constraints and false positives. These challenges are especially pronounced in the presence of heap aliasing, indirect calls, and extensive library or framework interactions, where precise modeling of program state becomes difficult~\cite{puasuareanu2019symbolic, pham2019enhancing}.

Dynamic constraint analysis techniques mitigate some precision limitations by deriving constraints from concrete executions, but introduce a different set of challenges. Because dynamic approaches rely on observed executions, they inherently provide incomplete coverage and may fail to discover constraints associated with unexecuted yet feasible paths. Their effectiveness depends heavily on input quality, test generation strategies, and user interactions, making comprehensive constraint discovery difficult to guarantee~\cite{godefroid2005dart, sen2006cute}. Moreover, dynamic constraint analysis does not provide soundness guarantees with respect to unobserved behaviors, as the absence of an observed execution does not imply infeasibility. These techniques can also incur substantial runtime overhead and exhibit nondeterministic behavior, particularly in concurrent or event-driven applications where execution traces may vary across runs~\cite{cadar2011symbolic}.

\subsection{Leveraging Static Analysis and LLM Reasoning}
To address the limitations of both static and dynamic constraint analysis paradigms, we propose a framework that combines static inference with LLM-assisted reasoning for backward constraint analysis. Static analysis provides structural coverage, while LLM reasoning is leveraged to interpret semantic information that is implicit in framework interactions, variable usage, and conditional logic. By restricting analysis to control-flow paths that are relevant to a specified target operation, the framework reduces spurious path exploration and improves precision. As execution paths are discovered, LLM-assisted semantic reasoning incrementally refines path conditions by interpreting conditional semantics and assembling precise reachability constraints across complex, event-driven execution flows. Rather than aiming for full semantic soundness or completeness, our approach prioritizes the discovery of human-interpretable reachability constraints that support practical security analysis tasks.

%Static approaches excel at broad coverage and soundness but often suffer from imprecision due to over-approximation. Dynamic approaches, in contrast, yield high-fidelity constraints grounded in concrete executions, but provide inherently incomplete coverage.

\section{Design of \tool}
This section presents the design of \tool—a backward constraint analysis framework for discovering and extracting human-interpretable reachability constraints from Android applications. The framework addresses the fundamental challenge of understanding what conditions must be satisfied to reach target operations in an unknown application. %Our approach combines the precision of static program analysis with large language model–assisted semantic reasoning to bridge the gap between low-level bytecode analysis and explicit, interpretable representations of targeted execution paths and their associated reachability constraints.

\begin{figure}[h]
    \centering
    \includegraphics[width=1.0\linewidth]{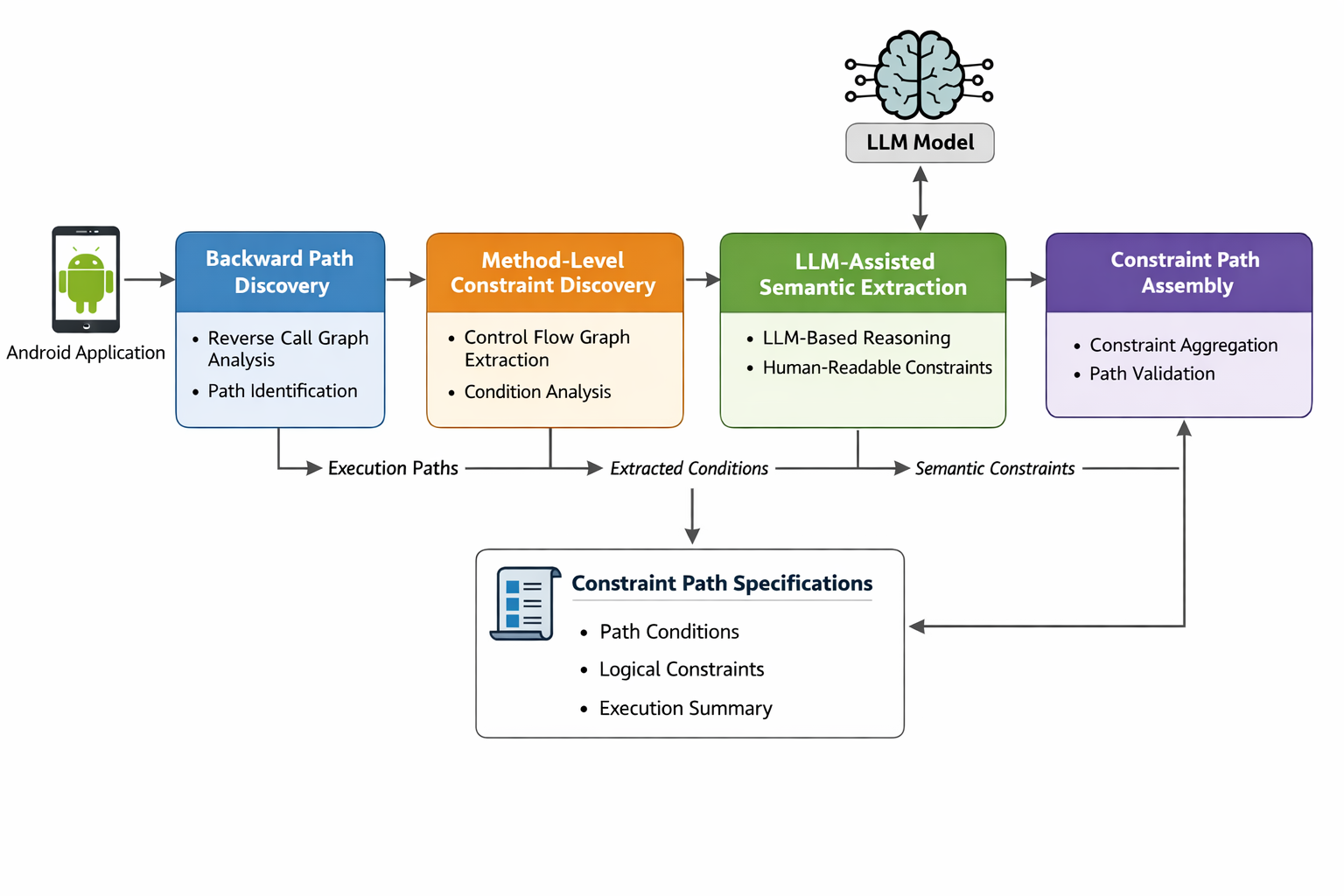}
    \caption{\tool Framework workflow}
    \label{fig:scope_workflow}
    \vspace{-0.5cm}
\end{figure}

Our framework consists of four main phases: 1) Backward Path Discovery; 2) Intraprocedural Constraint Discovery; 3) LLM-Assisted Semantic Extraction; and 4) Constraint Path Assembly. These phases work together sequentially to identify execution paths, extract control-flow conditions, interpret semantic meaning, and generate path constraint specifications, as shown in Figure~\ref{fig:scope_workflow}.
\vspace{-0.2cm}

\subsection{Backward Path Discovery}
In backward path discovery, we identify interprocedural execution paths from target methods to application entry points through call graph traversal. Let $T$ denote a target method within an Android application, and $G = (V, E)$ denote the application's call graph where $V$ is the set of reachable methods and $E$ represents caller-callee relationships. To accurately model Android lifecycle transitions and callback-driven execution—and thereby reduce spurious call edges—we construct $G$ using FlowDroid’s call graph~\cite{arzt2014flowdroid}. The backward path discovery phase then computes a set of execution paths $P = \{p_1, p_2, \ldots, p_n\}$ where each path $p_i$ is defined as:

\begin{equation}
p_i = \langle m_1, m_2, \ldots, m_k, T \rangle
\label{eq1}
\end{equation}

where:
\begin{itemize}
\item $T$ is the target method requiring constraint analysis; 
\item $m_1$ is an application entry point, such as an Android lifecycle callback, event handler, or broadcast receiver;
\item $\{m_2, \ldots, m_k\}$ are intermediate methods that form a valid interprocedural execution sequence;
\item Each consecutive method pair $(m_i, m_{i+1}) \in E$ corresponds to a caller–callee relationship in the call graph.
\end{itemize}

The backward path discovery algorithm, shown in Algorithm~\ref{alg:backward-path}, performs a breadth-first traversal over the reverse call graph. To ensure termination and scalability, the algorithm explicitly encodes the following mechanisms: (1) Cycle detection, where a method is expanded only if it does not already appear in the current path. This check is enforced in Algorithm~\ref{alg:backward-path}, line~9, by preventing revisits of methods already present in the active path. This mechanism prevents infinite exploration caused by recursion or cyclic call relationships. (2) Depth bounding, which limits exploration to paths of finite length using a configurable maximum depth parameter, denoted as $MAX\_DEPTH$. It is important to note that cycle detection is path-sensitive rather than global, allowing the same method to appear in multiple distinct execution paths while preventing infinite recursion within any single path. This design preserves backward-discovered candidate execution paths from application entry points to the target method. We note that the feasibility of \tool finding constraints on complete execution paths is relative to the precision of this static call graph.

\begin{algorithm}
\caption{Backward Path Discovery}
\label{alg:backward-path}
\begin{algorithmic}[1]
\Procedure{DiscoverPaths}{$T$, $G$, $MAX\_DEPTH$}
    \State Initialize queue $Q \leftarrow \{T\}$ and path set $P \leftarrow \emptyset$
    \While{$Q \neq \emptyset$ \textbf{and} depth $< MAX\_DEPTH$}
        \State Dequeue current method $m_i$ 
        \State Find all callers $C(m_i) = \{m_j \mid (m_j, m_i) \in E\}$
        \For{each caller $m_j \in C(m_i)$}
            \If{$m_j$ is an entry point}
                \State Add complete path to $P$
            \ElsIf{$m_j$ not in current path}
                \State Enqueue $m_j$
            \EndIf
        \EndFor
    \EndWhile
    \State \textbf{return} complete path set $P(T)$
\EndProcedure
\end{algorithmic}
\end{algorithm}

This target-driven backward analysis approach is chosen over forward exploration to focus the analysis on execution paths relevant to the target method. We avoid exploring large portions of the program that do not contribute to target reachability by initiating analysis from a target method and tracing its prerequisites backward through the call graph. As a result, our approach achieves significantly improved scalability compared to exhaustive forward symbolic execution, particularly in large, event-driven Android applications.

%We illustrate backward path discovery using the running example shown in Figure~\ref{fig:jimple_onoptionsitemselected} method \texttt{savePrice(String, float)}. Backward discovery begins at this method and traces caller relationships to identify application entry points from which it can be reached. Using the reverse call graph constructed by Algorithm~\ref{alg:backward-path}, \tool determines that \texttt{savePrice} is reachable via the \texttt{onOptionsItemSelected} callback in the \texttt{AddEditNoteActivity} class. The traversal then reaches the synthesized \texttt{dummyMainMethod}, indicating an application entry point. The resulting backward execution path is: \texttt{dummyMainMethod} $\rightarrow$ \texttt{AddEditNoteActivity.onOptionsItemSelected} $\rightarrow$ \texttt{AddEditNoteActivity.savePrice}. This path is subsequently passed to the next component for detailed intraprocedural constraint analysis.

\subsection{Intraprocedural Constraint Discovery}
For each execution path identified during backward path discovery, this phase analyzes all of its methods individually using an upward traversal algorithm to determine the reachability constraints that govern execution of the target method $T$. Given an discovered path in Equation \ref{eq1} $p_i = \langle m_1, \ldots, m_k, T \rangle$, we start with the immediate caller $m_k$ of the target method $T$. For this method, the analysis evaluates all conditional constructs that influence whether execution reaches $T$, aggregating the constraints required for invocation. The analysis then proceeds iteratively to each predecessor method along the path, continuing until all methods in $p_i$ have been analyzed. By restricting analysis to control-flow elements only along the candidate backward paths, the approach avoids the path explosion associated with forward symbolic execution while ensuring comprehensive coverage of constraints relevant to target reachability. 

We construct an intraprocedural control-flow graph for each method $m_i$, denoted as $CFG_{m_i}$, that captures only the control-flow elements relevant to constraint discovery. Unlike a traditional control-flow graph that includes all program statements, $CFG_{m_i}$ is a simplified representation that focuses on: (i) conditional statements that introduce branching behavior, (ii) switch statements representing multi-way decisions, (iii) method invocation sites, including calls to the target method, (iv) allocation sites relevant to variable definition, and (v) the control-flow edges connecting these elements.

Formally, we define $CFG_{m}$ as a directed graph $(D, E_i)$, where $D_i$ denotes decision nodes (IF\_NODE), switch nodes (SW\_NODE) and invocation nodes (INVOKE\_NODE), and $E_i$ represents control-flow relationships between these nodes. Each node encodes its corresponding Jimple statement along with the execution context required for subsequent constraint extraction.

Given $CFG_{m}$ and a target invocation site $callsiteT$ for $T$ , the upward traversal algorithm, as shown on Algorithm \ref{alg:upward_constraint}, initializes from $callsiteT$ and proceeds using a worklist-based exploration strategy. At each step, the algorithm examines only those predecessor nodes that can reach the current node in $CFG_{m}$, as returned by the \textsc{GetBranch} function, which returns only the predecessor IF\_NODEs and SW\_NODEs that can reach the current node $c$ in $CFG_m$. This reachability filtering ensures that analysis remains restricted to control-flow elements relevant to the target invocation and avoids exploring infeasible or unrelated paths. When the traversal encounters a decision point—IF\_NODEs or SW\_NODEs—the algorithm then invokes the LLM-assisted semantic interpretation component (explained in the next phase). Non-decision nodes, such as straight-line control-flow segments or allocation sites, are traversed without constraint extraction but are evaluated by the LLM for variable semantic analysis and framework interpretation. The traversal continues until all relevant predecessor nodes have been explored or the method entry point is reached. A visited set is maintained to prevent redundant analysis and ensure termination, even in the presence of cyclic intraprocedural control-flow structures.

%\color{red}Figure~\ref{fig:_onOptionItemSelected} \color{black}illustrates the  constructed for the \texttt{onOptionsItemSelected} method, which is a caller of \texttt{savePrice()} in our running example.

%As shown in Algorithm~\ref{alg:upward_constraint}, this phase performs intraprocedural upward traversal within each method in $CFG_{m_i}$, starting from the target method invocation and tracing backward to aggregate all conditions that must be satisfied for the invocation to occur. This analysis approach is applied iteratively across methods along the backward-discovered execution paths, enabling interprocedural constraint composition. 

\begin{algorithm}[h]
\caption{Upward Traversal for Intraprocedural Constraint Discovery}
\label{alg:upward_constraint}
\small
\begin{algorithmic}[1]
\Require $CFG_m$: simplified intraprocedural CFG for $m$
\Require $callsite_T$: target invocation node in $CFG_m$
\Require $entry_m$: method entry node in $CFG_m$
\Require $\kappa_m$: method bytecode/context (e.g., Jimple)
\Ensure $C$: reachability constraints required to reach $callsite_T$

\State $C \leftarrow \emptyset$
\State $W \leftarrow \{callsite_T\}$ \Comment{Worklist (stack/queue)}
\State $visited \leftarrow \emptyset$

\While{$W \neq \emptyset$}
    \State $c \leftarrow$ \Call{Pop}{$W$}
    \If{$c \in visited$}
        \State \textbf{continue}
    \EndIf
    \State $visited \leftarrow visited \cup \{c\}$

    \If{$c = entry_m$}
        \State \textbf{continue}
    \EndIf

    \State $Preds \leftarrow$ \Call{GetBranch}{$c, CFG_m$}

    \If{$Preds = \emptyset$}
        \State \textbf{continue}
    \EndIf

    \ForAll{$p \in Preds$}
        \If{$p$ is \textsc{IfNode} \textbf{or} $p$ is \textsc{SwitchNode}}
            \State $C \leftarrow \Phi_{\mathrm{LLM}}(p, \kappa_m, C)$
        \EndIf
        \State \Call{Push}{$W, p$}
    \EndFor
\EndWhile

\State \Return $C$
\end{algorithmic}
\end{algorithm}

\subsection{LLM-Assisted Semantic Analysis}\label{llmdesign}
Recent studies have demonstrated that large language models are capable of program summarization~\cite{ahmed2024automatic,ahmed2022few,zhang2025comprehensive}, code analysis~\cite{nam2024using, krishna2025codellm, blyth2025static, walton2024exploring}, and code generation~\cite{mundler2025type, guo2024stop}, suggesting that LLMs can capture semantic, syntactic, and control-flow relationships in code artifacts through carefully designed prompting strategies. Motivated by these findings, we leverage LLMs to interpret bytecode-level conditions that are difficult to reason about using purely syntactic or structural analysis.

\tool’s LLM-assisted semantic analysis is integrated directly into the intraprocedural upward constraint discovery phase, as shown in Algorithm~\ref{alg:upward_constraint}. As upward traversal progresses from the target invocation toward the method entry, the analysis visits decision points in reverse order. At each step, the LLM interprets the current decision node in the context of previously accumulated constraints, enabling incremental interpretation and refinement of reachability constraints as additional program context becomes available.

Let $C_i$ denote the evolving reachability-constraint state after processing decision node $d_i$. For each decision node, the LLM produces a candidate semantic interpretation:
\begin{equation}
\hat{r}_i = \Phi_{\mathrm{LLM}}(d_i, \kappa_i, C_{i-1}),
\label{eq:llm_semantic}
\end{equation}
where $\kappa_i$ represents the bytecode-level scope and contextual information associated with $d_i$, and $\Phi_{\mathrm{LLM}}$ infers the control-flow outcome required for target reachability, resolves variable semantics under the current program state, and interprets relevant framework or library method calls. The output $\hat{r}_i$ represents an LLM-generated candidate reachability constraint.

Our Semantic analysis in this phase is guided by a structured prompt design (Appendix~\ref{appendix:llm_prompt}) that directs the LLM through a sequence of well-defined reasoning tasks with explicit requirements on output structure. These tasks include:

\textbf{Branch Direction Analysis}: Inferring the control-flow outcome required for target reachability, including whether execution proceeds along the true or false branch of a conditional predicate or through a specific switch-case selection. This analysis leverages contextual information about conditional variables and execution flow that may not be evident from conditional syntax alone.

\textbf{Variable Semantic Analysis}: Resolving the semantic interpretation of variables appearing in conditional expressions. Bytecode-level analysis yields low-level variable identifiers that lack explicit semantic meaning. The LLM performs variable tracing and data-flow–aware interpretation, incorporating allocation sites and reaching definitions to characterize variable semantics under the current program state and within the conditional context. \emph{We note that this variable semantic analysis relies on available reaching definitions, not full interprocedural value resolution.}

\textbf{Framework Method Interpretation}: Interpreting the semantic behavior of framework and library method calls appearing within conditional expressions. Android applications extensively rely on framework APIs whose behavior significantly influences control flow but is difficult to infer from bytecode alone. The LLM leverages its exposure to large-scale code repositories to infer the semantics of such framework methods and their impact on execution flow.

\textbf{Constraint Reassessment}: Aggregating the results of branch direction analysis, variable semantic analysis, and framework method interpretation to update the accumulated constraint state. At each decision point, previously produced constraints are reconsidered in light of newly inferred semantics, enabling incremental refinement as traversal proceeds.

\paragraph{LLM Output Validation \& Hallucination Mitigation}
LLM-generated candidate interpretations $\hat{r}_i$ are treated as candidate semantic explanations and are accepted into the analysis only when consistent with evidence extracted by static analysis. Specifically, inferred branch outcomes are validated against control-flow reachability within the $CFG_m$, variable semantics are checked against def–use relationships and variable types, and framework method interpretations are constrained by known method signatures and invocation context. 

Thus, each candidate interpretation $\hat{r}_i$ is accepted only if it is consistent with the available structural evidence:

\[
 C=\hat{r}_i,  \text{IF } {Consistent}(\hat{r}_i, d_i, \kappa_i, C_{i-1}) == True\\
\]

Accepted interpretations update the constraint state $C_i$, while inconsistent interpretations are discarded or revised. This validation step ensures that incremental constraint refinement is grounded in extracted control-flow and data-flow evidence. This formulation checks for hallucination and prevents the LLM from introducing reachability conditions unsupported by the underlying program structure.

\begin{figure}[h]
\centering
\begin{lstlisting}[
    language=Java,
    basicstyle=\scriptsize\ttfamily,
    numbers=left,
    numberstyle=\tiny,
    frame=single,
    breaklines=true,
    showstringspaces=false,
    escapeinside={(*@}{@*)}
]
$i0 = interfaceinvoke $r1.<MenuItem: int getItemId()>();
$i1 = <R$id: int save_note>;
(*@\colorbox{red!20}{if \$i0 != \$i1 goto label3;}@*)
$r2 = virtualinvoke r0.<Intent getIntent()>();
$z0 = virtualinvoke $r2.<Intent: boolean hasExtra(String)>
      ("EXTRA_ID");
(*@\colorbox{orange!20}{if \$z0 == 0 goto label2;}@*)
$z0 = virtualinvoke $r2.<Intent: boolean hasExtra(String)>
      ("EXTRA_PRICE_CHECK"); 
(*@\colorbox{yellow!20}{if \$z0 == 0 goto label1;}@*)
$r3 = virtualinvoke $r2.<Intent: String getStringExtra(String)>
      ("EXTRA_TITLE");
$f0 = virtualinvoke $r2.<Intent: float getFloatExtra(String,float)>
      ("EXTRA_QUANTITY", 0.0F);
(*@\colorbox{green!20}{specialinvoke r0.<void savePrice(String,float)>(\$r3, \$f0);}@*)
\end{lstlisting}
\caption{Jimple bytecode from \texttt{onOptions()} showing conditional statements and the target invocation, with decision points represented as $D=\langle d_1,\ldots,d_n\rangle$.}
\label{fig:jimple_onoptionsitemselected}
\end{figure}

To illustrate intraprocedural constraint discovery with LLM-assisted semantic analysis, consider the bytecode for the \texttt{onOptionsItemSelected} method shown in Figure~\ref{fig:jimple_onoptionsitemselected}. Let $T$ denote the target callsite corresponding to the \texttt{savePrice(String, float)} invocation. During upward traversal of the method's $CFG_m$, the first decision point encountered is the conditional statement \texttt{if $z0 == 0 goto label1$} (line~10).
The LLM receives this conditional node together with the complete method bytecode and the target method \texttt{savePrice}. By analyzing the surrounding control-flow structure, the LLM determines that taking the true branch transfers execution to \texttt{label1}, which leads to an alternative code path rather than the target invocation. Consequently, reaching \texttt{savePrice} requires the condition to evaluate to false, yielding the local reachability constraint \texttt{$z0 != 0$}.
However, the low-level variable \texttt{$z0$} provides little semantic insight on its own. To resolve its meaning, the variable semantic analysis task traces \texttt{$z0$} to its most recent definition:
\texttt{$z0 = virtualinvoke $r2.<Intent: boolean hasExtra(String)>("EXTRA\_PRICE\_CHECK")} (line~8 in Figure~\ref{fig:jimple_onoptionsitemselected}). From this variable assignment, the LLM infers that \texttt{$z0$} represents the boolean result of checking whether an Android \texttt{Intent} contains the parameter \texttt{"EXTRA\_PRICE\_CHECK"}. The variable is therefore mapped to the human-readable concept \texttt{intentHasExtraPriceCheck}, reflecting its role in the application's execution logic.
To complete the interpretation, the framework method interpretation task evaluates the API call \texttt{Intent.hasExtra(String)}. The LLM recognizes this as an Android framework method that returns \texttt{true} if the specified key is present in the \texttt{Intent}'s data bundle and \texttt{false} otherwise. Combining this semantic understanding with the inferred branch outcome, the LLM refines the low-level constraint \texttt{$z0 != 0$} into the human-readable reachability requirement: the Intent must contain the \texttt{"EXTRA\_PRICE\_CHECK"} parameter.
After processing this decision point, the algorithm replaces the current node with its predecessor and continues the upward traversal to the next conditional statement. Applying the same semantic interpretation process to each encountered decision node, the intraprocedural analysis ultimately derives three boolean reachability constraints for the \texttt{savePrice} target:
\texttt{menuItemId == saveNoteId},
\texttt{intentHasExtraId == true}, and
\texttt{intentHasExtraPriceCheck == true}.

%Together, these constraints precisely characterize the conditions under which execution within \texttt{onOptionsItemSelected} can reach the target method \texttt{savePrice}.

Therefore the reachability constraint need to reach the \texttt{savePrice} from \texttt{onOptionsItemSelected} is:

\begin{align}
C(\text{savePrice}) = &\;(\text{menuItemId} == \text{saveNoteId}) \land \nonumber \\
                            &\;(\text{intentHasExtraId} == \text{true}) \land \nonumber \\
                            &\;(\text{intentHasExtraPriceCheck} == \text{true})
\end{align}
%This logical combination represents the complete set of execution requirements that must be satisfied for the discovered execution path to successfully reach the \texttt{savePrice} target method.

%when the condition evaluates to true, execution branches to \texttt{label1}, which leads to an alternative method invocation rather than the intended target. Conversely, when the condition evaluates to false, execution proceeds sequentially through the remaining bytecode and ultimately
%Each processed conditional produces a structured reachability constraint that preserves technical precision while remaining human-readable, enabling seamless integration with the final constraint path assembly component.

\subsection{Constraint Path Assembly}
The final phase assembles complete execution path requirements by conjunctively composing the finalized intraprocedural constraint sets across methods along a backward-discovered call path.

Given the backward-discovered execution path 
$p = \langle m_1, m_2, \ldots, m_k, T \rangle$, 
where each method $m_i$ contributes a set of intraprocedural reachability constraints $C_{m_i}$ required to reach the callsite to $m_{i+1}$ (or $T$ when $i = k$), the target-relevant path constraint for $p_i$ is:

\begin{equation}
C_p = \bigwedge_{i=1}^{k} \left( \bigwedge_{c \in C_{m_i}} c \right).
\end{equation}

Each $C_{m_i}$ represents a conjunction of necessary reachability predicates within method $m_i$; disjunctive alternatives within a method are not enumerated because our goal is to recover necessary reachability requirements rather than complete symbolic path formulas, trading exhaustive formal precision for scalability and human interpretability. Consequently, each $C_{m_i}$ summarizes the predicates required to reach the next callsite within ${m_i}$, and conjunctive composition across methods captures the joint preconditions for end-to-end reachability. Variable naming consistency is enforced during semantic extraction, enabling direct composition without additional interprocedural renaming. This design yields a scalable and interpretable approximation well suited for security analysis.

The output of this phase is a \textbf{Constraint Path Specification} that includes a \textbf{Unique Path Identifier} for tracking and reference purposes, the complete \textbf{Method Sequence} from application entry point to target method, the \textbf{Target-relevant path Constraint Expression} representing all execution requirements, and a \textbf{Natural Language Execution Summary} that describes the complete execution constraints in human readable terms.

%For each execution path discovered during backward traversal, the constraint assembly process phase aggregates the semantic constraints from all intermediate methods to produce a unified path specification that captures the complete execution prerequisites for reaching the target method.
%The constraints along an execution path follow logical AND semantics, as all intermediate conditions must be satisfied for execution to successfully reach the target method. For path $p_i$ with semantic constraints $S_i(p_i)$ extracted by the LLM analysis, the complete path constraint aggregated these individual requirements into a comprehensive specification. 

%\clearpage
\section{Implementation}

We implement \tool as an automated analysis framework for extracting reachability constraints from Android applications. The framework which has $\approx 7000$ lines of code is developed in Java and built on top of the Soot static-analysis framework~\cite{vallee1999soot}, enabling bytecode parsing and call-graph construction. Our implementation operates directly on Jimple, Soot’s typed three-address intermediate representation, which provides a structured and analyzable form of Dalvik bytecode. The pipeline accepts an Android APK file and target method(s) as inputs and produce structured reachability constraint specifications without requiring manual configuration or environment modeling.

The static-analysis component constructs an interprocedural call graph using FlowDroid \cite{arzt2014flowdroid} for each application and performs backward traversal from he input target method(s)s to identify candidate execution paths. For each discovered method on each path, \tool builds a CFG that selectively retains conditional statements, switch constructs, and invocation sites relevant to reachability. Extracted statements and their execution context are packaged in structured form and provided to the semantic-interpretation module, where constraint information is incrementally aggregated to produce complete execution-path specifications. 

The semantic-interpretation module integrates multiple large language models through a structured prompting interface. For each decision node, the framework invokes the integrated models to analyze the current context for reachability constraints. The LLM returns structured JSON outputs describing branch-direction requirements, variable semantics, and inferred constraints. To improve reliability, \tool includes a validation layer that checks generated constraints against control-flow reachability and def–use relationships derived from static analysis. All intermediate and final constraints are stored in structured JSON format to support automated aggregation and downstream analysis. The framework supports both local inference for open-source models and API-based inference for commercial models

\section{Evaluation}
%We evaluate \tool with respect to its core objectives—precision, interpretability, and scalability—through a series of empirical studies designed to assess both the quality of extracted reachability constraints and the practical performance of the framework. In particular, we seek to answer the following research questions:

We evaluate \tool with respect to its core objectives: accuracy, interpretability, and scalability, through a series of empirical studies designed to assess both the quality of extracted reachability constraints and the framework’s practical performance. In particular, we address the following research questions.

\begin{itemize}
    \item \textbf{RQ1}: How accurately can different LLMs perform task-specific semantic interpretation of Android framework APIs and bytecode-level conditions for reachability constraint extraction?
    \item \textbf{RQ2}: How does LLM-enhanced backward constraint analysis compare to traditional symbolic execution in terms of accuracy, performance, and interpretability?  
    \item \textbf{RQ3}: Can our framework effectively identify and explain security-relevant execution constraints in real-world malicious Android applications?
\end{itemize}

%\begin{table*}[t]
%\centering
%\small
%\caption{Overview of Selected Large Language Models}
%\label{tab:llm_overview}
%\footnotesize
%\begin{tabular}{|l|c|c|c|c|}
%\hline
%\textbf{Model} & \textbf{Architecture} & \textbf{Context} & %\textbf{Params} & \textbf{Release} \\
%\textbf{Name} & & \textbf{Length} & & \textbf{Date} \\
%\hline
%GPT-4o & GPT-4 & 128k & Undisclosed & May 2024 \\
%\hline
%DeepSeek-Coder & DeepSeek & 16k & 6.7B & June 2024 \\
%\hline
%Qwen-Coder & Qwen2 & 32k & 7B & April 2024 \\
%\hline
%CodeLlama & Llama2 & 16k & 7B & August 2023 \\
%\hline
%Llama-3 & Llama3 & 8k & 8B & April 2024 \\
%\hline
%\end{tabular}
%\end{table*}

%\textbf{Target Method Selection:} Given that our framework is designed to extract execution constraints through backward path analysis from target methods, we applied a systematic filtering strategy to include only target methods with sufficient complexity for constraint evaluation. From the original. containing conditional statements and associated target method invocations, performing static execution tracing to derive the reachability constraints required to reach each target.

\subsection{Experimental Setup}
\subsubsection{Benchmarks}
Existing Android analysis benchmarks, such as DroidBench~\cite{arzt2014flowdroid} and ICCBench~\cite{yan2022comprehensive}, primarily target taint analysis and data-flow tracking, making them unsuitable for evaluating constraint-extraction capabilities that require semantic understanding of execution prerequisites. To address this limitation, we construct a comprehensive test suite from real-world Android applications specifically designed to evaluate LLM performance on backward constraint analysis. We describe how we constructed the test suite in the following:

%\textbf{1. API-Only:}We constructed a corpus of 156 Android framework methods spanning six major API categories: Intent operations, menu item handling, activity lifecycle, UI component events, platform compatibility, and security APIs. These categories cover commonly used framework methods that introduce conditional control-flow dependencies in Android applications.
\textbf{1. API-Only:} We constructed a corpus of 156 Android framework methods spanning six major API categories: intent operations, menu item handling, activity lifecycle, UI component events, platform compatibility, and security APIs. These categories cover commonly used framework methods that introduce conditional control-flow dependencies in Android applications.
The ground-truth semantic descriptions for each framework method were manually curated by the research team through consultation of official Android documentation and validated via manual review.

%\textbf{2. Real-world Applications:} We selected 50 Android applications from F-Droid based on four critical criteria: (1) recent updates within 12 months to ensure modern Android development patterns, (2) presence of complex control flow patterns with at least three conditional branches or switch cases, (3) diverse Android framework method usage spanning multiple API categories, and (4) variable reuse patterns across different semantic contexts that challenge traditional static analysis approaches. This selection methodology ensures our evaluation covers the challenging constraint patterns encountered in production Android applications while maintaining reproducibility through open-source availability. From these 50 apps, we identified 78 target methods representing diverse Android interaction patterns, including menu handlers, Intent processors, activity lifecycle methods, authentication flows, platform compatibility checks, and security validation routines.
\textbf{2. Real-world applications:} We selected 50 Android applications from F-Droid \cite{fdroid} based on four criteria: (1) recent updates within the past 12 months to reflect modern development practices, (2) complex control-flow patterns with at least three conditional branches or switch cases, (3) diverse Android framework-method usage spanning multiple API categories, and (4) variable reuse across distinct semantic contexts that challenge traditional static analysis. This methodology ensures coverage of challenging constraint patterns observed in production Android applications while maintaining reproducibility through open-source availability. From these applications, we identified 78 target methods representing diverse interaction patterns, including menu handlers, intent processors, activity lifecycle methods, authentication flows, platform compatibility checks, and security validation routines. The ground-truth reachability constraints were constructed by the research team through static, call-graph–guided execution tracing over the Jimple bytecode of the selected applications. For each target method, tracing proceeded from the method entry point toward the target, analyzing statements sequentially to identify conditional checks along the execution path. At each conditional statement, the required branch outcome was determined and the corresponding reachability constraint was derived. This process continued until the target method was reached, yielding a conjunction of constraints that characterizes the execution requirements along the path.

\paragraph{Constraint Complexity.}
We further categorize test cases based on the number of constraints in the final specification and the presence of variable reuse across constraints:
\begin{itemize}
    \item \textbf{SIMPLE:} 0--3 constraints with no variable reuse across predicates (26 test cases)
    \item \textbf{MEDIUM:} 4--10 constraints with limited variable reuse across predicates (38 test cases)
    \item \textbf{COMPLEX:} More than 10 constraints and extensive variable reuse across predicates (14 test cases)
\end{itemize}

%\textbf{3. Malware:} we downloaded the most recent 100 Android malware samples from established malware repositories~\cite{allix2016androzoo, virusshare}. This benchmark specifically targets 42 dangerous Android APIs across 6 security categories.
\textbf{3. Malware:} We downloaded the 100 most recent Android malware samples from established malware repositories~\cite{allix2016androzoo, virusshare}. This benchmark targets 42 dangerous Android APIs spanning six security categories.

Our evaluation uses this benchmark to assess different aspects of \tool's constraint-extraction capability. For \textbf{RQ1}, we use the API-only dataset for framework-method interpretation and the real-world applications for constraint reasoning. For \textbf{RQ2}, we evaluate a representative subset of 15 applications for the real-world dataset (three per category across five major categories), ensuring balanced comparison while maintaining computational feasibility for symbolic execution. For \textbf{RQ3}, we leverage the malware dataset to evaluate the framework’s robustness under malicious scenarios.

\subsubsection{LLM Model Selection}
%To evaluate the effectiveness of the LLM in semantic analysis, we selected five representative LLMs based on architectural diversity, parameter scale, and practical availability for reproducible research. Our selection includes one commercial model (GPT-4o from OpenAI) and four open-source alternatives with 6.7-8 billion parameters: DeepSeek-Coder (6.7B), Qwen-Coder (7B), CodeLlama (7B), and Llama-3 (8B). This diversity enables comprehensive evaluation across different architectures, training methodologies, and resource requirements, as detailed in Table~\ref{tab:llm_overview}. All models received identical structured prompts containing complete method bytecode, specific conditional statements under analysis, target method, and comprehensive variable definition contexts. Prompts averaged 4,500 tokens and explicitly requested structured JSON output with dedicated sections for branch directions, constraint expressions, logic interpretations, and the constraint path specifications. Temperature parameters were uniformly set to 0.1 across all models to ensure deterministic constraint extraction and minimize response variability that could confound comparative analysis.

To evaluate the effectiveness of LLM-based semantic analysis, we selected five representative models based on architectural diversity, parameter scale, and practical availability for reproducible research. The set includes one commercial model (GPT-4o, OpenAI) and four open-source alternatives with 6.7–8B parameters: DeepSeek-Coder (6.7B), Qwen-Coder (7B), CodeLlama (7B), and Llama-3 (8B). This diversity enables evaluation across architectures, training paradigms, and resource requirements, as summarized in Table~\ref{tab:llm_overview}. All models received identical structured prompts containing full method bytecode, the conditional statement under analysis, the target method, and complete variable-definition context. Prompts averaged 4,500 tokens and requested structured JSON outputs specifying branch directions, constraint expressions, logical interpretations, and path-level constraint specifications. The temperature was fixed at 0.1 for all models to promote deterministic constraint extraction and minimize variability that could confound comparative analysis.

\begin{table}[t]
\centering
\footnotesize
\caption{Overview of Selected Large Language Models}
\label{tab:llm_overview}
\setlength{\tabcolsep}{3pt}
\begin{tabular}{lcccc}
\toprule
\textbf{Model} & \textbf{Arch.} & \textbf{Context} & \textbf{Params} & \textbf{Release} \\
               &                & \textbf{Length}  &                 & \textbf{Date}    \\
\midrule
GPT-4o         & GPT-4    & 128k & Undiscl. & May 2024 \\
DeepSeek-Coder & DeepSeek & 16k  & 6.7B     & Jun 2024 \\
Qwen-Coder     & Qwen2    & 32k  & 7B       & Apr 2024 \\
CodeLlama      & Llama2   & 16k  & 7B       & Aug 2023 \\
Llama-3        & Llama3   & 8k   & 8B       & Apr 2024 \\
\bottomrule
\end{tabular}
\end{table}

\subsubsection{Hardware Configuration}
%All experiments were conducted on a local server equipped with an RTX 4090 GPU (24GB VRAM), Intel i9-13900KF CPU (24 cores), and 64GB DDR5 RAM to ensure consistent performance conditions across all evaluations. Local open-source models were executed using this configuration with GPU acceleration enabled for transformer inference, while GPT-4o was accessed through OpenAI's API with standard rate limiting protocols. This setup enabled controlled, reproducible evaluation without external dependencies for open-source models while maintaining identical environmental conditions across all experimental runs.
All experiments were conducted on a local server equipped with an RTX 4090 GPU (24 GB VRAM), an Intel i9-13900KF CPU (24 cores), and 64 GB DDR5 RAM to ensure consistent performance across evaluations. Open-source models were executed locally with GPU acceleration for transformer inference, while GPT-4o was accessed via the OpenAI API under standard rate-limiting constraints. This configuration enabled controlled, reproducible evaluation for open-source models while maintaining identical environmental conditions across all experimental runs.

\subsection{Results}
\subsubsection{RQ1: Accuracy of LLM semantic Interpretation}
%To address RQ1, we evaluate LLM performance in two stages. 

\paragraph{RQ1a: Android Framework Method Semantic Interpretation.}
%We first assess framework method interpretation accuracy, as correct understanding of Android API semantics is a prerequisite for reachability constraint extraction and cannot be recovered through structural consistency checks alone. This evaluation focuses on the ability of LLMs to correctly identify and interpret Android framework methods when such calls appear within conditional statements. Specifically, we evaluate each model’s ability to (i) identify variable assignments originating from framework method invocations, (ii) describe the behavior of the invoked method, and (iii) specify possible return values and their semantic meaning.
We first assess framework-method interpretation accuracy, as correct understanding of Android API semantics is a prerequisite for reachability-constraint extraction and cannot be recovered through structural consistency checks alone. This evaluation examines the ability of LLMs to correctly identify and interpret Android framework methods appearing within conditional statements or decision node context. Specifically, we evaluate each model’s ability to (i) identify variable assignments originating from framework-method invocations, (ii) describe the invoked method’s behavior, and (iii) specify possible return values and their semantic meaning.

\textbf{Evaluation Methodology:}
LLM-generated framework method interpretations are compared against the 156 API-only manually curated ground truth descriptions. To account for paraphrased but semantically equivalent descriptions, we measure semantic similarity between the ground truth and LLM-generated descriptions using Sentence-BERT (SBERT) embeddings\cite{reimers2019sentence} with cosine similarity:

\begin{equation}
\text{similarity}(v_1, v_2) = \frac{v_1 \cdot v_2}{\|v_1\| \times \|v_2\|}
\end{equation}

where $v_1$ and $v_2$ denote the embedding vectors of the ground truth and LLM-generated descriptions, respectively. 

Since each framework invocation corresponds to a single semantic interpretation decision, we use accuracy as the primary evaluation metric rather than precision or recall. For each API category $C$, accuracy is computed as:

\begin{equation}
\text{Accuracy}(C) = \frac{\text{Number of methods with similarity} \geq T}{\text{Total methods in category } C}
\end{equation}

%An interpretation is considered correct if its accuracy score meets or exceeds a threshold $T = 0.6$. The similarity threshold of 0.6 is chosen to tolerate lexical variation while preserving semantic equivalence, consistent with prior work on LLM-based semantic evaluation.
An interpretation is considered correct if its accuracy score meets or exceeds the threshold $T = 0.6$. This similarity threshold tolerates lexical variation while preserving semantic equivalence, consistent with prior work on LLM-based semantic evaluation.

%\textbf{RQ1a Results:} GPT-4o achieved the highest framework method interpretation accuracy across all evaluated API categories, exceeding 90\% accuracy in five of the six categories, as shown in Table~\ref{tab:framework_accuracy}. The model consistently produced semantically aligned interpretations that accurately reflected framework method behavior and return semantics. For example, when analyzing \texttt{Intent.hasExtra("EXTRA\_ID")}, GPT-4o correctly identified the method as checking for the presence of a specific parameter in an \texttt{Intent} and returning a boolean value indicating whether the parameter exists—behavior consistent with the Android framework specification. The performance gap between GPT-4o and open-source models was most pronounced for \textbf{Platform Compatibility APIs} category. These APIs often require awareness of Android version evolution and API deprecation patterns, which may be underrepresented in general-purpose code corpora. Notable disparities were also observed for \textbf{Security APIs}.

\textbf{RQ1a Results:} GPT-4o achieved the highest framework-method interpretation accuracy across all evaluated API categories, exceeding 90\% accuracy in five of the six categories (Table~\ref{tab:framework_accuracy}). The model consistently produced semantically aligned interpretations that reflected framework-method behavior and return semantics. For example, when analyzing \texttt{Intent.hasExtra("EXTRA\_ID")}, GPT-4o correctly identified the method as checking for the presence of a specific parameter in an \texttt{Intent} and returning a boolean indicating whether the parameter exists, consistent with the Android framework specification. The largest performance gap between GPT-4o and open-source models occurred for \textit{Platform Compatibility APIs}, which often require awareness of Android version evolution and API deprecation patterns that may be underrepresented in general-purpose code corpora. Notable disparities were also observed for \textit{Security APIs}. \vspace{-0.8cm}

%likely reflecting the specialized and less frequently encountered nature of authentication and security-related framework methods.

\begin{table*}[h]
\centering
\caption{Android Framework Method Modeling Accuracy by API Category}
\label{tab:framework_accuracy}
\footnotesize
\begin{tabular}{|l|c|l|c|c|c|c|c|}
\hline
\textbf{API Category} & \textbf{Method} & \textbf{Example Methods} & \textbf{GPT-4o} & \textbf{DeepSeek} & \textbf{Qwen} & \textbf{CodeLlama} & \textbf{Llama-3} \\
 & \textbf{Count} &  & \textbf{(\%)} & \textbf{(\%)} & \textbf{(\%)} & \textbf{(\%)} & \textbf{(\%)} \\
\hline
\textbf{Intent Operations} & 42 & hasExtra(), getStringExtra(), setFlags() & 95.2 & 67.3 & 73.8 & 52.4 & 58.1 \\
\hline
\textbf{Menu Item Handling} & 28 & getItemId(), setVisible(), findItem() & 96.4 & 71.4 & 78.6 & 60.7 & 64.3 \\
\hline
\textbf{Activity Lifecycle} & 24 & finish(), getIntent(), onBackPressed() & 91.7 & 58.3 & 66.7 & 45.8 & 50.0 \\
\hline
\textbf{UI Component Events} & 18 & onClick(), setOnClickListener(), findViewById() & 94.4 & 66.7 & 72.2 & 55.6 & 61.1 \\
\hline
\textbf{Platform Compatibility} & 22 & SDK\_INT, getOnBackPressedDispatcher() & 86.4 & 45.5 & 54.5 & 31.8 & 36.4 \\
\hline
\textbf{Security APIs} & 22 & authenticate(), secureDialog(), checkNotNull() & 90.9 & 54.5 & 63.6 & 40.9 & 45.5 \\
\hline
\textbf{Overall Average} & \textbf{156} & -- & \textbf{92.6} & \textbf{62.2} & \textbf{68.6} & \textbf{49.4} & \textbf{54.5} \\
\hline
\end{tabular}
\begin{tablenotes}
\item \textit{Note: Accuracy measured using semantic embeddings (SBERT) with threshold of 0.6 between LLM-generated framework method descriptions and validated ground truth interpretations.}
\end{tablenotes}
\end{table*}

\begin{table*}[h]
\centering
\caption{Overall Constraint Extraction Performance by Complexity Category}
\label{tab:constraint_performance}
\footnotesize
\begin{tabular}{|l|c|c|c|c|c|c|}
\hline
\textbf{Test Case Category} & \textbf{Total} & \textbf{GPT-4o} & \textbf{DeepSeek-Coder} & \textbf{Qwen-Coder} & \textbf{CodeLlama} & \textbf{Llama-3} \\
 & \textbf{Cases} & \textbf{P/R/F1} & \textbf{P/R/F1} & \textbf{P/R/F1} & \textbf{P/R/F1} & \textbf{P/R/F1} \\
\hline
\textbf{SIMPLE} & 26 & 96.2/94.2/95.2 & 78.3/72.1/75.1 & 82.7/76.9/79.7 & 71.4/65.4/68.3 & 73.1/67.3/70.1 \\
\hline
\textbf{MEDIUM} & 38 & 91.6/87.3/89.4 & 69.4/61.2/65.1 & 74.2/68.1/71.0 & 58.7/52.3/55.3 & 61.3/55.8/58.4 \\
\hline
\textbf{COMPLEX} & 14 & 88.2/83.6/85.8 & 52.4/45.7/48.9 & 67.3/58.9/62.8 & 41.2/35.1/37.9 & 44.6/38.2/41.2 \\
\hline
\textbf{Overall} & \textbf{78} & \textbf{92.3/88.7/90.4} & \textbf{69.1/61.4/65.0} & \textbf{75.2/68.3/71.6} & \textbf{58.4/52.1/55.1} & \textbf{61.0/55.2/58.0} \\
\hline
\end{tabular}
\begin{tablenotes}
\item \textit{Note: P = Precision, R = Recall, F1 = F1-Score (\%). Precision measures correctly identified constraints versus total generated; Recall measures identified constraints versus ground truth requirements.}
\end{tablenotes}
\end{table*}

%\paragraph{RQ1b: Constraint Reasoning: }
%As the second component of RQ1, we evaluate end-to-end constraint reasoning on 50 real-world Android applications comprising 78 target methods. It is important to distinguish between structural consistency checks in \tool design and correctness as measured in RQ1b. Consistency checks ensure that LLM-generated interpretations do not contradict extracted control-flow or data-flow evidence and therefore prevent hallucination and  infeasible reachability conditions. However, multiple semantically distinct interpretations may remain structurally consistent. The RQ1b evaluation measures whether the framework identifies the correct reachability constraints among these structurally valid candidates, relative to manually curated ground truth. The assessment encompasses the full constraint extraction pipeline, including interpretation of conditional statements, framework method semantics, variable context resolution, and synthesis of human-readable constraint specifications. We evaluate all five LLMs across the full set of 78 target methods, with test cases categorized by complexity to analyze performance degradation as constraint complexity increases.

\paragraph{RQ1b: Constraint Reasoning.}
Next, we evaluate end-to-end constraint reasoning on 50 real-world Android applications comprising 78 target methods. It is important to distinguish structural consistency checks in \tool from correctness as measured in RQ1b. Consistency checks ensure that LLM-generated interpretations do not contradict extracted control-flow or data-flow evidence, thereby preventing hallucinations and infeasible reachability conditions. However, multiple semantically distinct interpretations may remain structurally consistent. RQ1b therefore evaluates whether the framework identifies the correct reachability constraints among these structurally valid candidates relative to manually curated ground truth. The assessment covers the full constraint-extraction pipeline, including interpretation of conditional statements, framework-method semantics, variable-context resolution, and synthesis of human-readable constraint specifications. We evaluate all five LLMs across the 78 target methods, categorizing test cases by complexity to analyze performance degradation as constraint complexity increases.

\textbf{Evaluation Methodology:} A generated constraint set is considered \emph{completely correct} if: (i) the number of generated constraints matches the ground truth specification, and (ii) each generated constraint is logically equivalent to its corresponding ground truth constraint. Partial matches are not considered successful.

For each complexity category $C$, accuracy is computed as:
\begin{equation}
\text{Accuracy}_C = \frac{\text{\# of test cases w/ complete correct constraint sets}}{\text{Total test cases in category } C}
\end{equation}
This strict accuracy metric ensures the resulting output from LLM produce accurate, end-to-end constraint specifications. To further provide a more granular view of constraint quality, we additionally compute precision, recall, and F1-score across all test cases:
\begin{align}
\text{Precision} &= \frac{\text{Number of correctly identified constraints}}{\text{Total generated constraints}} \\
\text{Recall} &= \frac{\text{Number of correctly identified constraints}}{\text{Total required constraints}} \\
\text{F1-score} &= \frac{2 \times \text{Precision} \times \text{Recall}}{\text{Precision} + \text{Recall}}
\end{align}

A constraint is considered correctly identified when it matches the corresponding ground truth constraint in both logical structure and semantic interpretation. We report both strict accuracy and precision/recall metrics to distinguish complete constraint synthesis failures from partial semantic understanding.

%\textbf{RQ1b Results.}
%The evaluation results show that GPT-4o consistently outperformed all other evaluated models across all complexity categories, achieving an overall F1-score of 90.4\%, as summarized in Table~\ref{tab:constraint_performance}. The performance gap is most pronounced in \textbf{COMPLEX} scenarios, where GPT-4o attained an F1-score of 85.8\%. These cases involve dense and interdependent conditional logic, platform compatibility checks, and security-related validation constraints. DeepSeek-Coder and Qwen-Coder exhibited moderate performance, achieving overall F1-scores of 65.0\% and 71.6\%, respectively. In contrast, CodeLlama and Llama-3 demonstrated more limited constraint reasoning capabilities, with F1-scores below 60\%. Manual inspection of failure cases indicates that open-source models struggled primarily with accurate interpretation of Android framework method semantics (further validating RQ1a) and with complex variable reuse patterns, where the same bytecode variables represent distinct semantic concepts across multiple conditional statements. 

\textbf{RQ1b Results.} The evaluation results show that GPT-4o consistently outperformed all other evaluated models across all complexity categories, achieving an overall F1-score of 90.4\% (Table~\ref{tab:constraint_performance}). The performance gap is most pronounced in \textbf{Complex} scenarios, where GPT-4o attained an F1-score of 85.8\%. These cases involve dense, interdependent conditional logic, platform compatibility checks, and security-related validation constraints. DeepSeek-Coder and Qwen-Coder exhibited moderate performance, with overall F1-scores of 65.0\% and 71.6\%, respectively. In contrast, CodeLlama and Llama-3 showed more limited constraint-reasoning capability, with F1-scores below 60\%. Manual inspection of failure cases indicates that open-source models struggled primarily with accurate interpretation of Android framework-method semantics, further supporting RQ1a, and with complex variable-reuse patterns in which the same bytecode variables represent distinct semantic concepts across multiple conditional statements. Beyond absolute performance metrics, consistency of constraint extraction across varying complexity levels is critical for practical analysis workflows. Figure~\ref{fig:llm_consistency} in Appendix~\ref{appendix:llmconsistency} presents a consistency analysis revealing substantial reliability differences across evaluated LLMs. GPT-4o demonstrates not only the highest overall precision but also the most stable performance across complexity categories, with precision ranging from 88.2\% to 96.2\%—a narrow 8-percentage-point span. This stability indicates robust constraint-extraction performance even as execution paths and constraint interactions grow more complex. In contrast, open-source models exhibit significantly higher variance with increasing complexity. CodeLlama shows the largest variability, with precision ranging from 41.2\% to 71.4\%, while DeepSeek-Coder and Llama-3 display similarly wide fluctuations. Such variability makes constraint-extraction reliability difficult to predict across target methods, posing challenges for consistent analyst use in real-world settings.

%These results highlight the importance of strong semantic reasoning capabilities for reliable constraint extraction from Android bytecode. In particular, the increasing variance observed for open-source models as complexity grows underscores the role of advanced semantic interpretation in bridging the gap between low-level bytecode analysis and human-interpretable constraint specifications.

\begin{tcolorbox}[left=2pt,right=2pt,top=2pt,bottom=2pt,colback=gray!5,colframe=gray!50,boxrule=0.5pt,title={\small RQ1 Key Findings}]
\small
Results show that stronger LLMs (notably GPT-4o) achieve high accuracy and robust performance, particularly for complex, multi-constraint execution paths, while weaker models degrade significantly as semantic complexity increases. These results highlight the importance of strong semantic reasoning capabilities for reliable constraint extraction
from Android bytecode. In particular, the increasing variance observed for open-source models as complexity grows
underscores the role of advanced semantic interpretation in
bridging the gap between low-level bytecode analysis and
human-interpretable constraint specifications.
\end{tcolorbox}

%subsubsection{RQ2: Comparative Analysis %with Traditional Symbolic Execution}
%Having established that GPT-4o achieves the strongest overall performance across the metrics evaluated in RQ1, we integrate this model in the final version of \tool and conduct a comparative evaluation against traditional symbolic execution approaches. Using 15 applications from our F-Droid benchmark subset, we compare \tool against angr~\cite{shoshitaishvili2016state}, a widely used symbolic execution framework.

%This evaluation addresses a fundamental question in program analysis: while LLM-assisted techniques can reason about program semantics, do they offer measurable advantages over established symbolic execution methods in terms of scalability and practical utility, while preserving precision in constraint discovery? We assess this through three complementary metrics: RQ2a: analysis time to evaluate performance efficiency (scalability), RQ2b: constraint equivalence relative to ground truth to assess precision, and RQ2c: interpretability to measure the usefulness of extracted constraints for security analysis workflows.

\subsubsection{RQ2: Comparative Analysis with Traditional Symbolic Execution}
Given GPT-4o’s strong overall performance across the RQ1 metrics, we integrate this model into the final version of \tool and conduct a comparative evaluation against traditional symbolic execution. Using 15 applications from our F-Droid benchmark subset, we compare \tool with angr~\cite{shoshitaishvili2016state}, a widely used symbolic-execution framework.

This evaluation addresses a central question in program analysis: although LLM-assisted techniques can reason about program semantics, do they provide measurable advantages over established symbolic-execution methods in scalability and practical utility while preserving practical precision in reachability constraint discovery? We assess this using three complementary metrics: \textbf{RQ2a}, analysis time as a measure of scalability; \textbf{RQ2b}, constraint equivalence relative to ground truth as a measure of precision; and \textbf{RQ2c}, interpretability for security-analysis workflows.

%\paragraph{RQ2a: Analysis Time: }
%ur framework operates as an end-to-end analysis pipeline that constructs call graphs, performs backward path traversal from target methods to application entry points, and incrementally generates semantic reachability constraints within a single execution. The reported analysis time therefore reflects a deployment-oriented workflow in which an analyst specifies a target method and receives human-interpretable constraints without intermediate configuration steps. In contrast, traditional symbolic execution tools typically require substantial upfront configuration for Android applications. This configuration commonly includes manual identification of entry points, symbolic state initialization for Android framework components, and integration of appropriate SDK and bytecode analysis support.

\textbf{RQ2a Results.} Our framework operates as an end-to-end analysis pipeline that constructs call graphs, performs backward traversal from target methods to application entry points, and incrementally generates semantic reachability constraints within a single execution. The reported analysis time therefore reflects a deployment-oriented workflow in which an analyst specifies a target method and receives human-interpretable constraints without intermediate configuration. In contrast, traditional symbolic-execution tools typically require substantial upfront configuration for Android applications, including manual identification of entry points, symbolic state initialization for Android framework components, and integration of appropriate SDK and bytecode-analysis support.

%To account for these differing operational models, our reported analysis time for \tool is the complete workflow from target method specification to constraint generation. For angr, we measure constraint discovery execution time after completing the necessary setup phases, including Android SDK integration and symbolic state configuration. A timeout threshold of 30 minutes per target method is enforced to bound symbolic execution path exploration within practical limits. Across the evaluation dataset, our approach demonstrates consistent performance advantages, achieving a 5.8$\times$ faster average analysis time and successfully completing analysis for 100\% of target methods, compared to an 80\% completion rate for symbolic execution, as shown in Table~\ref{tab:comparative_analysis}. Symbolic execution timeouts occurred predominantly in applications with extensive Android framework interactions or complex control-flow structures, indicative of the well-known scalability challenges when applying traditional symbolic execution to real-world, framework-intensive Android applications.

To account for these differing operational models, the reported analysis time for \tool covers the complete workflow from target-method specification to constraint generation. For angr, we measure constraint-discovery time after completing the required setup phases, including Android SDK integration and symbolic-state configuration. A timeout of 30 minutes per target method bounds symbolic-execution path exploration within practical limits. Across the evaluation dataset, our approach demonstrates consistent performance advantages, achieving a 5.8$\times$ faster average analysis time and completing analysis for 100\% of target methods, compared to an 80\% completion rate for symbolic execution, as shown in Table~\ref{tab:comparative_analysis}. Timeouts occur primarily in applications with extensive Android framework interactions or complex control-flow structures, reflecting known scalability challenges of applying traditional symbolic execution to real-world, framework-intensive Android applications.

%\begin{table*}[h]
%\centering
%\caption{Constraint Discovery Performance Comparison}
%\label{tab:comparative_analysis}
%\footnotesize
%\begin{tabular}{|l|l|c|c|c|}
%\hline
%\textbf{App Category} & \textbf{Target Method Type} & \textbf{\tool} & \textbf{Angr} & \textbf{Constraint} \\
% &  & \textbf{(seconds)} & \textbf{(seconds)} & \textbf{Equivalence} \\
%\hline
%Productivity & Menu Event Handler & 24 & 114 & \checkmark \\
%\hline
%Financial & SMS Verification & 31 & 287 & \checkmark \\
%\hline
%Social Media & Location Access & 45 & —* & N/A \\
%\hline
%Gaming & Network Request & 18 & 156 & \checkmark \\
%%\hline
%Utility & Permission Check & 22 & 98 & \checkmark \\
%\hline
%\textbf{Average} &  & \textbf{28.0} & \textbf{163.8} & \textbf{80\%} \\
%\hline
%\textbf{Success Rate} &  & \textbf{100\%} & \textbf{80\%} &  \\
%\hline
%\end{tabular}
%\begin{tablenotes}
%\item \textit{* Analysis timeout exceeded 30-minute threshold}
%\end{tablenotes}
%\end{table*}

\begin{table}[t]
\centering
\footnotesize
\caption{Constraint Discovery Performance Comparison}
\label{tab:comparative_analysis}
\setlength{\tabcolsep}{3pt}
\begin{tabular}{l l c c c}
\toprule
\textbf{App Category} & \textbf{Target Method} & \textbf{\tool} & \textbf{Angr} & \textbf{Constraint} \\
                      & \textbf{Type}          & \textbf{(sec)} & \textbf{(sec)} & \textbf{Equivalence} \\
\midrule
Productivity  & Menu Event Handler & 24 & 114 & \checkmark \\
Financial     & SMS Verification   & 31 & 287 & \checkmark \\
Social Media  & Location Access    & 45 & ---* & N/A \\
Gaming        & Network Request    & 18 & 156 & \checkmark \\
Utility       & Permission Check   & 22 & 98  & \checkmark \\
\midrule
\textbf{Average}      &            & \textbf{28.0}  & \textbf{163.8} & \textbf{80\%} \\
\textbf{Success Rate} &            & \textbf{100\%} & \textbf{80\%}  &               \\
\bottomrule
\end{tabular}

\vspace{2pt}
\footnotesize{\textit{* Analysis timeout exceeded the 30-minute threshold.}}
\end{table}

\paragraph{RQ2b: Constraint Equivalence.}
For applications where both approaches successfully completed analysis, we observe logical equivalence between the constraint sets produced by \tool and by angr. In these cases, both systems identify the same logical prerequisites required to reach the target methods, despite differences in analysis strategy and representation. This equivalence provides validation that the LLM-assisted semantic extraction in \tool preserves the reachability constraints identified by traditional symbolic execution, while expressing them in a more interpretable, analyst-oriented form. These results demonstrate that our approach maintains semantic alignment with established analysis techniques, supporting its use as a practical alternative for constraint discovery in real-world Android applications.
\paragraph{RQ2c: Interpretability Analysis.}
Differences in constraint representation are particularly relevant for practical analysis workflows. Using the Bazarnote productivity application as an illustrative case study, traditional symbolic execution produces constraints such as:

  \texttt{unc\_int\_getItemId\_0\_32 == default\_value\_int\_1\_32}, \texttt{unc\_boolean\_hasExtra\_3\_32 != 0x0}, and \texttt{unc\_boolean\_hasExtra\_4\_32 != 0x0}  

In contrast, \tool generates semantically equivalent but human-interpretable constraints: \texttt{menuItemId == saveNoteItemId}, \texttt{!intentHasExtraId}, and \texttt{intentHasExtraPriceCheck == true}. Interpreting the symbolic execution output requires additional manual reasoning to infer that the first constraint corresponds to menu item selection validation, while the remaining predicates check for specific \texttt{Intent} parameter states. By contrast, the constraints generated by \tool explicitly encode these semantics, directly conveying the execution requirements: the user must select a specific menu item, the \texttt{Intent} must not contain a particular identifier, and a price validation check must be enabled. This semantic clarity reduces interpretation effort and facilitates integration into security analysis workflows.

\begin{tcolorbox}[left=2pt,right=2pt,top=2pt,bottom=2pt,colback=gray!5,colframe=gray!50,boxrule=0.5pt,title={\small RQ2 Key Findings}]
\small
RECON achieves higher completion rates while producing constraint sets that are logically equivalent to those generated by angr when both analyses terminate. Additionally, RECON provides significantly more interpretable constraint representations, reducing analyst effort and improving practical usability without sacrificing semantic alignment.
\end{tcolorbox}

\subsubsection{RQ3: Malware Constraint Extraction Capability}
%Using our malware test cases, we evaluate whether \tool can be effectively applied to security analysis in real-world adversarial settings. In particular, we assess whether the framework can identify execution constraints leading to the invocation of security-sensitive (dangerous) Android APIs across multiple malware families.
Using our malware test cases, we evaluate whether \tool can be effectively applied to security analysis in real-world adversarial settings. Specifically, we assess whether the framework can identify execution constraints leading to the invocation of security-sensitive Android APIs across multiple malware families.

\textbf{Key Terms.}
\emph{Dangerous API identification} refers to detecting occurrences of security-sensitive framework method calls within malware code. \emph{Constraint extraction} denotes the generation of human-readable execution requirements for identified APIs. \emph{Completion rate} measures the percentage of identified API instances for which the framework successfully produces constraint specifications.

%As shown in Table~\ref{tab:malware_results}, \tool identifies dangerous API invocations across all six security categories represented in the malware dataset. Table~\ref{tab:api_distribution} shows that network data exfiltration APIs are the most prevalent, appearing in 78\% of analyzed samples, reflecting the need for outbound communication in many malware families. In contrast, privacy sensor APIs, while less frequent (34\% of samples), exhibit the highest average constraint complexity, with 8.9 conditions per execution path, indicating more restrictive and layered execution requirements for accessing sensitive device data.

As shown in Table~\ref{tab:malware_results}, \tool identifies dangerous API invocations across all six security categories in the malware dataset. Table~\ref{tab:api_distribution} further shows that network data-exfiltration APIs are the most prevalent, appearing in 78\% of analyzed samples and reflecting the reliance of many malware families on outbound communication. In contrast, privacy-sensor APIs, while less frequent (34\% of samples), exhibit the highest average constraint complexity at 8.9 conditions per execution path, indicating more restrictive and layered execution requirements for accessing sensitive device data. Overall, \tool achieves an 84\% constraint-extraction completion rate, successfully generating semantic reachability constraints for 325 of 387 identified dangerous API instances. This result indicates robustness in the presence of real-world malware behaviors and obfuscation. Failures occur primarily in samples employing heavy code obfuscation or complex reflection-based execution patterns, which restrict static control-flow and data-flow recovery.

%\begin{table}[h]
%\centering
%\caption{Malware Constraint Extraction Results}
%\label{tab:malware_results}
%\footnotesize
%\begin{tabular}{|l|c|}
%\hline
%\textbf{Metric} & \textbf{Value} \\
%\hline
%Malware Samples Analyzed & 100 \\
%\hline
%Total Dangerous API Types Found & 387 \\
%\hline
%Average APIs per Sample & 3.9 (Range: 1--8) \\
%\hline
%Constraint Extraction Success Rate & 325/387 (84\%) \\
%\hline
%Average Constraint Complexity & 6.2 conditions/path \\
%\hline
%\end{tabular}
%\end{table}

%\begin{table}[h]
%\centering
%\caption{API Category Distribution and Complexity}
%\label{tab:api_distribution}
%\footnotesize
%\begin{tabular}{|l|c|c|c|}
%\hline
%\textbf{API Category} & \textbf{Samples} & %\textbf{Instances} & \textbf{Avg} \\
%& \textbf{(\%)} & & \textbf{Complex.} \\
%\hline
%Network Exfiltration & 78\% & 145 & 4.3 \\
%\hline
%Device Info & 67\% & 98 & 7.1 \\
%\hline
%SMS/Phone & 45\% & 76 & 5.8 \\
%\hline
%Privacy Sensors & 34\% & 42 & 8.9 \\
%\hline
%File System & 23\% & 19 & 3.2 \\
%\hline
%Dynamic Loading & 12\% & 7 & 4.6 \\
%\hline
%\end{tabular}
%\end{table}

\begin{table}[t]
\centering
\caption{Malware Constraint Extraction Results}
\label{tab:malware_results}
\footnotesize
\begin{tabular}{l c}
\specialrule{1pt}{0pt}{0pt}
\textbf{Metric} & \textbf{Value} \\
\specialrule{1pt}{0pt}{0pt}
Malware Samples Analyzed & 100 \\
Total Dangerous API Types Found & 387 \\
Average APIs per Sample & 3.9 (Range: 1--8) \\
Constraint Extraction Success Rate & 325/387 (84\%) \\
Total Number of Paths & 1154\\
Total Number of Expression & 14472\\
Average Constraint Complexity & 6.2 conditions/path \\
\specialrule{1pt}{0pt}{0pt}
\end{tabular}
\end{table}

\begin{table}[t]
\centering
\caption{API Category Distribution and Complexity}
\label{tab:api_distribution}
\footnotesize
\begin{tabular}{l c c c}
\specialrule{1pt}{0pt}{0pt}
\textbf{API Category} & \textbf{Samples} & \textbf{Instances} & \textbf{Avg} \\
                      & \textbf{(\%)}    &                    & \textbf{Complex.} \\
\specialrule{1pt}{0pt}{0pt}
Network Exfiltration & 78\% & 145 & 4.3 \\
Device Info          & 67\% & 98  & 7.1 \\
SMS/Phone            & 45\% & 76  & 5.8 \\
Privacy Sensors      & 34\% & 42  & 8.9 \\
File System          & 23\% & 19  & 3.2 \\
Dynamic Loading      & 12\% & 7   & 4.6 \\
\specialrule{1pt}{0pt}{0pt}
\end{tabular}
\end{table}

%\textbf{Constraint Complexity Demonstration: }
%Figure~\ref{fig:malware_path_example} presents an illustrative execution path discovered during malware analysis, corresponding to one of 53 distinct paths leading to the \texttt{getLastKnownLocation()} API within a single malware sample. The path illustrates location access triggered during application termination (\texttt{onDestroy}) through integration with the Kiip advertising SDK, followed by background thread execution that ultimately invokes location data collection.

\textbf{Constraint Complexity Demonstration.} Figure~\ref{fig:malware_path_example} in Appendix~\ref{appendix:malware-use-case} presents an illustrative execution path discovered during malware analysis, corresponding to one of 53 distinct paths leading to the \texttt{getLastKnownLocation()} API within a single malware sample. The path illustrates location access triggered during application termination (\texttt{onDestroy}) through integration with the Kiip advertising SDK, followed by background thread execution that ultimately invokes location data collection. Further exploration of this malware use case is provided in the Appendix~\ref{appendix:malware-use-case}

\begin{tcolorbox}[left=2pt,right=2pt,top=2pt,bottom=2pt,colback=gray!5,colframe=gray!50,boxrule=0.5pt,title={\small RQ3 Key Findings}]
\small
RECON can identify and interpret execution constraints for dangerous API invocations across diverse malware families. Additionally, the constraint complexity case study illustrate RECON ’s ability to translate complex, multi-stage execution dependencies into human-
readable constraints, supporting its applicability to practical
malware analysis.
\end{tcolorbox}

\subsection{Limitations and Future Work}

First, our evaluation reveals a clear performance gap between proprietary LLMs (notably GPT-4o) and open-source alternatives. This dependence raises practical concerns regarding deployment cost, model availability, and long-term reproducibility as LLM ecosystems evolve. Although \tool is model-agnostic by design, its performance is currently limited by the semantic-reasoning capability of the underlying model. Second, \tool does not provide formal guarantees of soundness or completeness. Similar to prior static-analysis and symbolic-execution techniques, the framework aims to recover plausible and practically useful execution constraints rather than exhaustively enumerate all feasible paths. Third, \tool relies on recoverable control-flow and data-flow information. As observed in RQ3, malware employing heavy obfuscation, dynamic reflection, or runtime code loading can restrict static path recovery and reduce constraint-extraction coverage. This limitation is inherent to static-analysis–based approaches and is shared by traditional symbolic-execution frameworks. Finally, although our evaluation includes multiple real-world applications and malware families, the dataset size remains limited relative to the diversity of Android applications in the wild. Broader evaluation on larger and more diverse datasets may reveal additional edge cases and performance variability.

\noindent \textbf{Future Work} will focus on improving framework-method interpretation for open-source LLMs through lightweight fine-tuning or domain-adaptive instruction tuning on Android-specific APIs. Enhanced API semantic understanding is expected to improve downstream constraint-reasoning accuracy and narrow the performance gap observed in RQ1 and RQ2. We also plan to integrate execution paths and reachability constraints recovered by \tool with in-memory analysis and runtime artifact extraction. Combining static constraint reasoning with memory-resident state information may enable partial dynamic validation of execution paths, improving coverage in the presence of reflection or dynamically constructed control flows.
\section{Related Work}
\textbf{Constraints Reasoning and Symbolic Execution} :Prior work has investigated multiple approaches for extracting and reasoning about execution constraints, with symbolic execution emerging as one of the dominant techniques. Despite its effectiveness, symbolic execution faces well-known scalability challenges. The path explosion problem causes the number of possible execution paths to grow exponentially as programs encounter branching conditions, quickly exceeding available analysis resources.Several techniques attempt to mitigate path explosion through path prioritization or heuristic state selection \cite{yao2025empc,he2021learning}, but these strategies typically rely on manually designed heuristics that do not generalize across diverse application domains. Constraint solving also limits symbolic execution scalability. Complex path conditions involving nonlinear arithmetic, library calls, or environmental dependencies remain difficult for traditional solvers to handle efficiently. Researchers have explored alternative solving strategies, including heuristic search \cite{dinges2014solving}, machine-learning–guided constraint solving \cite{li2016symbolic,bu2021machine}, and constraint reuse or normalization techniques \cite{jia2015enhancing,brennan2017constraint}. Additional symbolic execution extensions, such as probabilistic symbolic execution \cite{borges2015iterative} and counterfactual symbolic execution \cite{hallahan2019lazy}, improve reasoning capabilities but still rely heavily on precise constraint modeling.

Android applications rely heavily on platform APIs whose behavior cannot be fully captured through static modeling alone. Prior work demonstrates that accurate framework modeling is essential for preserving constraint correctness during symbolic execution \cite{gao2018android,luo2017system}. Similar semantic challenges arise in specialized domains such as secure enclave analysis \cite{wang2023symgx} and cryptographic protocol verification \cite{aizatulin2012computational,chau2019analyzing}. Compilation-based and binary-level symbolic execution techniques improve execution efficiency \cite{255310,poeplau2021symqemu,daniel2020binsec}, but they continue to struggle when program behavior depends on high-level semantic understanding or obfuscated control flows \cite{yadegari2015symbolic}.

\textbf{AI-Enhanced Constraint Reasoning}: Recent research integrates machine learning and language models into program analysis workflows to improve constraint reasoning and path discovery. Learning-guided symbolic execution uses trained policies to prioritize symbolic states and improve coverage efficiency \cite{he2021learning,he2019learning}. State-aware symbolic execution further enhances analysis of programs with state-dependent behaviors, such as finite-state systems \cite{277220}. Machine learning has also been applied to assist constraint solving by identifying feasible solution regions for complex path conditions \cite{li2016symbolic,bu2021machine}.

Large language models have recently been explored as semantic reasoning engines for program analysis and security applications. Prior studies demonstrate that LLMs can generate semantic summaries and assist vulnerability discovery, improving analyst understanding of program behavior \cite{walton2024exploring}. However, recent benchmarking studies show that LLMs still struggle to accurately reason about program execution paths and path-specific constraints when generating targeted test inputs or coverage-driven test cases \cite{wang2025testeval}. These findings indicate that while LLMs offer strong semantic reasoning capabilities, effectively integrating them into structured constraint extraction workflows remains an open research challenge.

\textbf{AI in Malware and Android Security Analysis}: Artificial intelligence has been increasingly applied to malware detection and Android security analysis~\cite{li2024revisiting, li2025foredroid}. Some researchers uses neural approaches to learn behavioral patterns directly from dynamic execution reports, reducing reliance on manual feature engineering and improving generalization across malware families \cite{jindal2019neurlux}. Network traffic and behavioral modeling techniques have also been used to detect advanced malware communication channels \cite{dodia2022exposing}. Despite these advances, Android security analysis remains challenging due to the platform’s event-driven architecture and heavy reliance on framework APIs \cite{ruggia2024unmasking}.

\section{Conclusion}
Existing techniques for finding execution constraints, particularly symbolic execution, provide precise reasoning but struggle to scale to complex program due to path explosion, framework interactions, and limited interpretability of low-level program representations. To address these challenges, we presented RECON, an LLM-enhanced backward constraint analysis framework that integrates static program analysis with semantic reasoning to extract human-interpretable execution constraints from Android bytecode. Our evaluation demonstrates that RECON improves constraint discovery scalability and usability while preserving logical correctness. In real-world applications, RECON achieves a 5.8x perfromance improvement over traditional symbolic execution. Additionally, RECON achieves an 84\% constraint extraction success rate across 100 malware samples, showcasing its robustness to handle even complex programs.
\bibliographystyle{plainurl}
\bibliography{BIB}
\balance

%-------------------------------------------------------------------------------
% Mandatory appendices for USENIX Security
%-------------------------------------------------------------------------------

\appendices

\section{LLM Performance Consistency Analysis}
\label{appendix:llmconsistency}
\begin{figure}[h]
\centering
\includegraphics[width=0.4\textwidth]{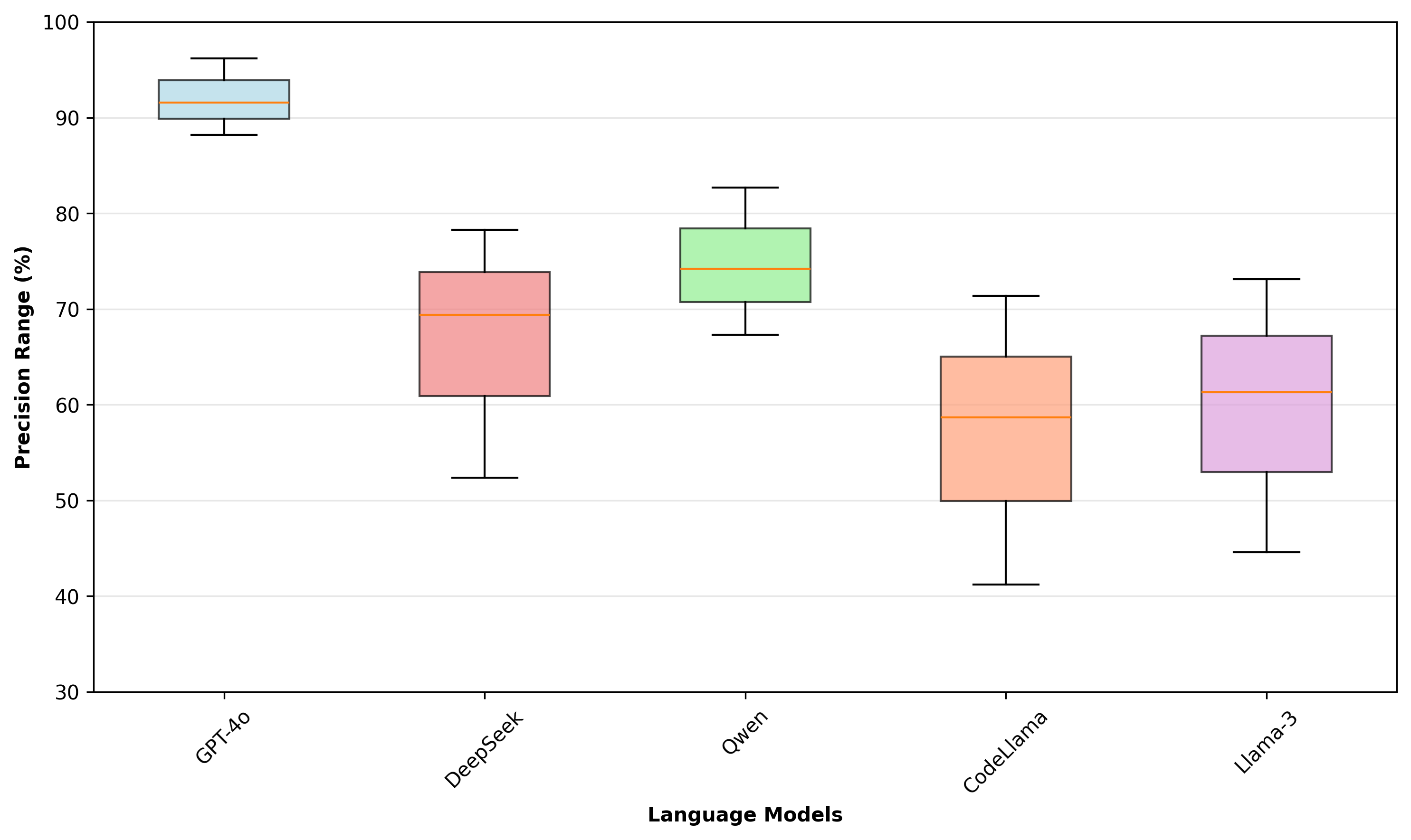}
\caption{LLM Performance Consistency Analysis. Box plot showing precision score distributions across complexity categories (SIMPLE, MEDIUM, COMPLEX). Box height indicates performance variability, while position shows overall performance level.}
\label{fig:llm_consistency}
\end{figure}

\section{RECON's LLM Prompt Template}
\label{appendix:llm_prompt}
\begin{figure}[H]
\centering
\small
\begin{tcolorbox}[left=4pt,right=4pt,top=4pt,bottom=4pt,colback=gray!5,colframe=gray!50,boxrule=0.5pt]
\textbf{1. Task:} You are an expert static analysis engineer specializing in Android bytecode constraint extraction. Analyze conditional statements in Android Jimple bytecode to determine execution constraints for target method reachability.
\tcblower
\textbf{2. Input:} \texttt{\{\{conditional\_node\}\}} --- Conditional node from CFG \\
\texttt{\{\{target\_node\}\}} --- Target node or Target method callsite \\
\texttt{\{\{method\_bytecode\}\}} --- Complete Jimple bytecode of containing method
\end{tcolorbox}
\begin{tcolorbox}[left=4pt,right=4pt,top=4pt,bottom=4pt,colback=gray!5,colframe=gray!50,boxrule=0.5pt]
\textbf{3. Instructions:} (1) Trace variables to recent assignments, (2) Interpret framework method semantics, (3) Determine branch direction (TRUE/FALSE/CASE\_X), (4) Replace bytecode names with semantic equivalents, (5) Generate boolean logic and human-readable constraints
\tcblower
\textbf{4. Output:} Provide response in this exact JSON format: \\
\texttt{\{} \\
\texttt{~~"branch\_direction": "TRUE|FALSE|CASE\_X",} \\
\texttt{~~"variable\_mappings": \{"original\_name": "semantic\_name"\},} \\
\texttt{~~"framework\_methods": [\{"method": "name", "purpose": "function"\}],} \\
\texttt{~~"boolean\_constraint": "semantic\_variable == value",} \\
\texttt{~~"business\_constraint": "Human readable requirement"} \\
\texttt{\}}
\end{tcolorbox}
\caption{LLM prompt template for semantic constraint extraction using structured JSON output.}
\label{fig:constraint_prompt_template}
\end{figure}

%\cleardoublepage

\section{Illustrative Malware Constraint-Complexity Use Case}
\label{appendix:malware-use-case}
Reaching the getLastKnownLocation() API requires satisfaction of eight semantic constraints, including the initialization state of the advertising SDK (\texttt{mKiipInitialized == true}), specific permission configurations (coarse location granted while fine location denied), and system service availability checks. These constraints capture layered execution dependencies that complicate straightforward static analysis.

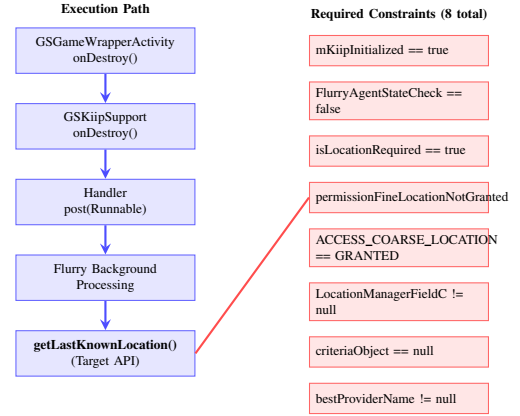
\begin{figure}[h]
\centering
\scriptsize
\begin{tikzpicture}[
    node distance=1.0cm,
    auto,
    pathbox/.style={rectangle, draw=blue!60, fill=blue!10, text width=2.2cm, text centered, minimum height=0.6cm, font=\tiny},
    constraintbox/.style={rectangle, draw=red!60, fill=red!10, text width=2.2cm, align=left, minimum height=0.4cm, font=\tiny},
    arrow/.style={thick, ->, >=stealth, blue!70},
    connector/.style={thick, -, red!70}
]
% Execution Path (Left Side) - closer spacing
\node [pathbox] (start) {GSGameWrapperActivity\\onDestroy()};
\node [pathbox, below of=start] (kiip) {GSKiipSupport\\onDestroy()};
\node [pathbox, below of=kiip] (handler) {Handler\\post(Runnable)};
\node [pathbox, below of=handler] (flurry) {Flurry Background\\Processing};
\node [pathbox, below of=flurry] (location) {\textbf{getLastKnownLocation()}\\(Target API)};

% Execution Path Arrows
\draw [arrow] (start) -- (kiip);
\draw [arrow] (kiip) -- (handler);
\draw [arrow] (handler) -- (flurry);
\draw [arrow] (flurry) -- (location);

% Constraints (Right Side) - much closer to left side
\node [constraintbox, right=1.5cm of start] (c1) {mKiipInitialized == true};
\node [constraintbox, below=0.2cm of c1] (c2) {FlurryAgentStateCheck == false};
\node [constraintbox, below=0.2cm of c2] (c3) {isLocationRequired == true};
\node [constraintbox, below=0.2cm of c3] (c4) {permissionFineLocationNotGranted};
\node [constraintbox, below=0.2cm of c4] (c5) {ACCESS\_COARSE\_LOCATION == GRANTED};
\node [constraintbox, below=0.2cm of c5] (c6) {LocationManagerFieldC != null};
\node [constraintbox, below=0.2cm of c6] (c7) {criteriaObject == null};
\node [constraintbox, below=0.2cm of c7] (c8) {bestProviderName != null};

% Connecting line from path to constraints
\draw [connector] (location.east) -- (c4.west);

% Labels
\node [above=0.1cm of start, font=\tiny\bfseries] {Execution Path};
\node [above=0.1cm of c1, font=\tiny\bfseries] {Required Constraints (8 total)};

\end{tikzpicture}
\caption{Malware location access execution path and constraint requirements. Shows one of 53 discovered paths to getLastKnownLocation() API, demonstrating location access through advertising SDK integration during application termination.}
\label{fig:malware_path_example}
\end{figure}

The semantic constraints shown in Figure~\ref{fig:malware_path_example} highlight execution patterns in which malware leverages legitimate third-party frameworks and application lifecycle transitions to access sensitive device information. By translating these dependencies into human-interpretable execution requirements, \tool enables security analysts to quickly understand the conditions under which sensitive behaviors are triggered, without requiring extensive manual reverse engineering.

%%%%%%%%%%%%%%%%%%%%%%%%%%%%%%%%%%%%%%%%%%%%%%%%%%%%%%%%%%%%%%%%%%%%%%%%%%%%%%%%
\end{document}